\documentstyle[aps,psfig,amssymb,multicol]{revtex}
\newcommand{\be}{\begin{equation}}
\newcommand{\ee}{\end{equation}}
\newcommand{\br}{\begin{eqnarray}}
\newcommand{\er}{\end{eqnarray}}

\begin{document}

\title{Entanglement versus mixedness for coupled qubits under a
phase damping channel}
\author{E. S. Cardoso\footnote{esc@ifi.unicamp.br}, M. C. de Oliveira\footnote{marcos@ifi.unicamp.br},
K. Furuya\footnote{furuya@ifi.unicamp.br}}
\address{ Instituto de F\'\i sica Gleb Wataghin,  Universidade Estadual
de Campinas, 13083-970, Campinas - SP, Brazil.}
\date{\today}
%\vspace{2cm}
%\begin{figure}
%\centerline{$\;$\hskip 0 truecm\epsfig{figure=uq.eps,height=2 cm}}
%\end{figure}
\maketitle

\begin{abstract}
 Quantification of entanglement against mixing is given for a
 system of coupled qubits under a phase damping channel. A family
 of pure initial joint states is defined, ranging from pure
 separable states to maximally entangled state. An
 ordering of entanglement measures is given
 for well defined initial state amount of entanglement.
 %{\it Key words}: entanglement; quantum evolution; non-classical states,
%quantum information.

PACS numbers: 03.67.-a  03.67.Mn 03.65.Ta
\end{abstract}

%\begin{multicols}{2}
%\narrowtext \tighten
\section{Introduction}
Quantum entanglement is an essential resource for many quantum
communication protocols, such as teleportation
\cite{bennett,zeilinger} and dense coding
\cite{bennett2,zeilinger2}, allowing an information processing
efficiency, which is otherwise unattainable through classical
protocols. As such the efficiency of those quantum protocols
relies on the ability to isolate the encoding system, being
strongly decreased when the system is coupled to an external
environment, attaining thus the classical limits, when the quantum
system is found in a separable state \cite{nielsen}. The relation
of entanglement against separability for mixed entangled states
has generated a considerable literature, with many proposals for
quantifying it (see for example
\cite{Munro01,peres,horodecki,TzuChieh03,peters} and references
therein). Bipartite quantum systems with Hilbert space of
dimension $ 2\otimes 2$ (coupled qubits) have been exhaustively
investigated in order to achieve a precise quantification of
entanglement against mixing. Particularly a valuable necessary and
sufficient condition for separability of coupled qubits has been
given by Peres \cite{peres} and Horodecki \cite{horodecki} in
terms of the {\it positivity} of the partial transpose of the
system density matrix, which sets a boundary for comparison to
many proposed entanglement measures, such as entanglement of
formation, relative entropy of distillation, and relative entropy
of entanglement. A very useful paper has appeared recently
\cite{TzuChieh03} relating the ordering of many entanglement
measures in relation to the degree of mixedness. This paper
reinforce and extends the discussion presented in  Ref.
\cite{Munro01} on maximally entangled mixed states (MEMS), which
are states that for a given mixedness achieve the greatest
possible entanglement.
More recently the imbalance between the sensitivity of common
state measures, such as fidelity trace distances, concurrence,
tangle and von Neumann entropies when acted on by a depolarizing
channel have also been investigated \cite{peters}. It was noticed
that the size of the imbalance depends intrinsically of the state
tangle and of the state purity. The results of Refs.
\cite{TzuChieh03,peters} were derived for arbitrary entangled
mixed states - randomly generated bipartite matrices respecting
the structure of positive semidefinite operators.  An actual
bipartite interacting system has its entanglement constrained by
its dynamics and thus (for certain given initial states) they
never reach the discussed MEMS. The system's dynamics constrains
the degree of entanglement and mixedness to a bounded range. Only
the MEMS dynamically connected to the system state are important
for setting reference states, and thus only those states are valid
for entanglement quantification in terms of distance measures.
Those MEMS can be computed by a certain combination of the reduced
density matrix of the coupled qubits. It is thus of central
importance to analyze the amount of entanglement against mixing
present in a quantum system due to a process of deterministic
entanglement formation \cite{entfor} in a noisy channel.

In this paper we analyze the degree of entanglement against mixing
for a dynamical system composed of two coupled qubits under the
phase damping channel. While amplitude damping is certainly the
most important source of noise for light field states qubit
encoding, the phase damping model describes more appropriately
noise over an encoding system composed of internal atomic (ionic)
states or even for internal quantum dots states
\cite{CiracZoller95,CMonroe95,Sasura02,SchmidtKaler03,Murao98}.
The phase damping channel is particularly interesting in analyzing
the degree of entanglement against mixing because it truly induces
decoherence without amplitude relaxation effects \cite{nielsen}.
We compare many entanglement measures as a function of the joint
state purity and discuss how do they relate to each other for the
specific dynamical system considered. More specifically we compare
concurrence and negativity with Bures distance entanglement
measures. While concurrence and negativity are able to quantify
the amount of entanglement present in a mixed state, they are not
able to distinguish states. The Bures distance entanglement
measure, on the other hand, is able to distinguish states and thus
can be used to define a ordering of entanglement measures. In its
definition however, a deep analysis must be made on the reference
state, from where the distance measure is taken. We have
considered both mixed and pure reference states, since both
situations allows inspection of many distinct entanglement
features. In Sec. II we present the model of coupled qubits under
the phase damping channel. In Sec. III we present the measures
used to infer the amount of entanglement and mixing through the
paper. In Sec. IV we define entanglement measures in terms of
Bures distance to maximally mixed and maximally entangled states.
In Sec. V we compare all the discussed entanglement measures
against mixedness and comment a possible measure ordering for the
proposed model. Finally, Sec. VI concludes the paper.

\section{coupled qubits under phase damping channel}

The system considered is constituted of two qubits coupled by a
phase interaction under the effect of a common environment
constituted by $N$ harmonic oscillators. Since we are interested
in the effect of the environment over the purity of the coupled
qubit system, we consider a number conserving interaction to the
harmonic oscillator set. A straightforward application of the
procedures developed here is envisaged for qubits encoded in
internal states of trapped ions in the proposal of Cirac and
Zoller \cite{CiracZoller95,CMonroe95,Sasura02,SchmidtKaler03} for
quantum computation.
 Following the DiVincenzo criteria \cite{DiVincenzo}, quantum gate
operations must be shorter than the decoherence time.
 Past results have shown that decoherence in trapped ions appears
without energy exchange \cite{Murao98} and thus cannot be
explained by the physical processes that take in consideration the
energy exchange as a source of decoherence, and thus the
inadequacy of the amplitude damping model. Our model Hamiltonian
writes
\begin{eqnarray}
    H=H_{S}+H_{R}+H_{I},
    \end{eqnarray}
with
\begin{eqnarray}
    H_{S}=\omega_{1}S_{1z}+\omega_{2}S_{2z}+\mu_{12} S_{1z}S_{2z},
    \end{eqnarray}
\begin{eqnarray}
    H_{R}=\hbar \omega_{1}\sum^{N}_{k=1}\widetilde{\omega}_{k}\left(n_{k}+\frac{1}{2}\right)+2\hbar\mu\sum^{N}_{i<j}n_{i}n_{j}.
    \end{eqnarray}
\begin{eqnarray}
    H_{I}=\sum^{N}_{k=1}n_{k}\left(\mu_{1}S_{1z}+\mu_{2}S_{2z}\right),
\end{eqnarray}
with
\begin{eqnarray}
    \widetilde{\omega}_{k}=\frac{\omega_{k}}{\omega_{1}}\nonumber
\end{eqnarray}
In this model the reservoir, together with the proposed
interaction is responsible by decoherence without energy damping.

We shall investigate how the proposed model describes the
evolution from a pure maximally entangled state to a separable
state. For that we base our discussion on some entanglement
 measures  previously discussed \cite{Munro01,TzuChieh03,Verstraete01}.
Firstly consider  that the two qubit states are prepared in an
entangled pure state in contact to a reservoir prepared
 as such
\begin{eqnarray}
|\psi\left(0\right)\rangle&=&\frac{1}{\sqrt{2\varepsilon(\varepsilon
-1) + 1}} (\varepsilon |+,+\rangle+
(1-\varepsilon)|-,-\rangle)\nonumber\\
&&\otimes \prod^{N}_{i=1}|\alpha_{i}\rangle . \label{inst}
\end{eqnarray} The degree of entanglement of the initial state is
a function of $\varepsilon$, which varies from 0 to 1. The
maximally entangled pure initial state is reached for
$\varepsilon=0.5$, while pure separable states are obtained for
$\varepsilon=0$ and $1$. The evolved joint state given by the
evolution of the state (\ref{inst}) due to the Hamiltonian (1)-(4)
reads
%\end{multicols}
%\begin{widetext}
\begin{eqnarray}
    \rho(t)=\frac{1}{(2\varepsilon(\varepsilon -1) + 1)}\sum_{n_{1}\dots
n_{N}}\sum_{n^{'}_{1}\dots n^{'}_{N}}
\prod^{N}_{j=1}\frac{\alpha^{n_{j}}_{j}{\alpha^{*}}^{{n}^{'}_{j}}}{\sqrt{n_{j}!
n^{'}_{j}!}}\{ \varepsilon^2
e^{-i\{\phi_{1j}(t)-\phi^{'}_{1j}(t)\}} \mid +,+,n_{1},\dots
,n_{N}\rangle \langle +,+,n^{'}_{1},\dots, n^{'}_{N}\mid+
\nonumber\\
+(1-\varepsilon)^2 e^{-i\{\phi_{2j}(t)-\phi^{'}_{2j}(t)\}} \mid -,-,n_{1},\dots ,
n_{N}\rangle \langle -,-,n^{'}_{1},\dots, n^{'}_{N}\mid+\nonumber\\
+\varepsilon(1-\varepsilon)
e^{-i(\omega_{1}+\omega_{2})t}e^{-i\{\phi_{1j}(t)
-\phi^{'}_{2j}(t)\}}\mid +,+,n_{1},\dots ,n_{N}\rangle \langle
-,-,n^{'}_{1},
\dots, n^{'}_{N}\mid+\nonumber\\
+\varepsilon(1-\varepsilon)
e^{+i(\omega_{1}+\omega_{2})t}e^{+i\{\phi^{'}_{1j}(t)
-\phi_{2j}(t)\}}\mid -,-,n_{1},\dots ,n_{N}\rangle \langle
+,+,n^{'}_{1},\dots, n^{'}_{N}\mid\}\end{eqnarray}
%\end{widetext}
%\begin{multicols}{2}
with
\begin{eqnarray}
\phi_{1j}(t)=+n_{j}(\frac{\mu_{1}+\mu_{2}}{2})t+\omega_{1}\widetilde{\omega}_{j}t(n_{j}+\frac{1}{2})\nonumber\\+2\mu
\omega_{1}t
\sum_{k>j}\widetilde{\omega}_{k}n_{k}\widetilde{\omega}_{j}n_{j}
\end{eqnarray}
\begin{eqnarray}
\phi_{2j}(t)=-n_{j}(\frac{\mu_{1}+\mu_{2}}{2})t+\omega_{1}\widetilde{\omega}_{j}t(n_{j}+\frac{1}{2})\nonumber\\+2\mu
\omega_{1}t
\sum_{k>j}\widetilde{\omega}_{k}n_{k}\widetilde{\omega}_{j}n_{j}
\end{eqnarray}
and
\begin{eqnarray}
    \phi^{'}_{1j}(t)=\phi_{1j}(n \longrightarrow n^{'})\nonumber\\
    \phi^{'}_{2j}(t)=\phi_{2j}(n \longrightarrow n^{'}).
\end{eqnarray}
When we consider a particular case of a resonant bath, where all
the $\omega_j$'s of the bath are the same, one can obtain a closed
expression for the reduced density matrix of the coupled qubits,
which then writes as
\begin{eqnarray}\label{state}
\rho_{\varepsilon}=\left( \begin{array}{cccc}
\frac{\varepsilon^2}{(2\varepsilon (\varepsilon -1) + 1)} & 0 & 0
& \frac{\varepsilon(1-\varepsilon)}{(2
\varepsilon(\varepsilon -1) + 1)}A\\
0 & 0 & 0 & 0\\ 0 & 0 & 0 & 0\\ \frac{\varepsilon(1-\varepsilon)}{(2
\varepsilon(\varepsilon -1) + 1)}A^{*} & 0 & 0 & \frac{
(1-\varepsilon)^2}{(2 \varepsilon(\varepsilon -1) + 1)}
\end{array} \right),
\end{eqnarray} which clearly has the structure of a mixed
nonmaximally mixed state with
\begin{eqnarray}\label{a}
A&=&e^{-i(\omega_{1}+\omega_{2})t}e^{\sum^{N}_{j=1}\mid\alpha_{j}
\mid^{2}[e^{-i(\mu_{1}+\mu_{2})t}-1]}\nonumber\\
&=&e^{-i(\omega_{1}+\omega_{2})t}e^{\widetilde{N}[e^{-i(\mu_{1}+\mu_{2})t}-1]},
\end{eqnarray} and $\widetilde{N}\equiv \sum^{N}_{j=1}|\alpha_{j}|^{2}$. We then identify the typical operation sum structure of the
phase damping
channel\be\rho_{\varepsilon}=E_0\rho_{\varepsilon}^{0}E_0^\dagger+E_1\rho_{\varepsilon}^{0}E_1^\dagger,\ee
with \be E_0=\left(\begin{array}{cccc} 1 & 0 & 0 & 0\\
0 & 0 & 0 & 0\\ 0 & 0 & 0 & 0\\ 0 & 0 & 0 & \sqrt{1-\gamma}
\end{array} \right)\;\;\;E_1=\left(\begin{array}{cccc} 0& 0 & 0 & 0\\
0 & 0 & 0 & 0\\ 0 & 0 & 0 & 0\\ 0 & 0 & 0 & \sqrt{\gamma}
\end{array} \right),\ee and
\be \label{rho0}\rho_{\varepsilon}^{0}=\left(\begin{array}{cccc}
\frac{\varepsilon^2}{(2\varepsilon (\varepsilon -1) + 1)} & 0 & 0
&
\frac{\varepsilon(1-\varepsilon)e^{-i\phi(t)}}{(2\varepsilon (\varepsilon -1) + 1)} \\
0 & 0 & 0 & 0\\ 0 & 0 & 0 & 0\\
\frac{\varepsilon(1-\varepsilon)e^{i\phi(t)}}{(2\varepsilon
(\varepsilon -1) + 1)} & 0 & 0
&\frac{(1-\varepsilon)^2}{(2\varepsilon (\varepsilon -1) + 1)}
\end{array}\right),\ee
{where} $\gamma\equiv
1-e^{2\widetilde{N}[\cos{(\mu_{1}+\mu_{2})t}-1]}=1-|A|^2$, and
$e^{-i\phi(t)}\equiv e^{-i[
(\omega_1+\omega_2)t+\widetilde{N}\sin{(\mu_1+\mu_2)t}]}=A/|A|$.

\section{Entanglement measures and mixing}

We shall refer  to the state degree of mixing through the linear
entropy
\be\delta_{12}=\frac{d}{d-1}\left[1-\mbox{Tr}(\rho_{\varepsilon}^2)\right],
\ee where $d$ is the dimension of the system Hilbert space. The
linear entropy ranges from 0 (for pure states) to 1 (for maximally
mixed states $\rho_{MM}={I_{d}}/{d}$).  It has been determined
that arbitrary bipartite states whose linear entropy
$\delta_{12}\ge d(d-2)/(d-1)^2$ are separable \cite{TzuChieh03}.
For the coupled qubits system that means that states whose
$\delta_{12}\ge 8/9$ are certainly separable \cite{TzuChieh03}.
For the state considered here the linear entropy explicitly reads
\begin{eqnarray}
\delta_{12}(\varepsilon,t)&=&\frac{4}{3}\left\{\frac{2\varepsilon^2(1-\varepsilon)^2(1-|A|^2)}
{[\varepsilon^2+(1-\varepsilon)^2]^2}\right\}\nonumber\\
&=&\frac{4}{3}\left\{\frac{2\varepsilon^2(1-\varepsilon)^2\gamma}
{[\varepsilon^2+(1-\varepsilon)^2]^2}\right\}.
\end{eqnarray}
Since $0\le\varepsilon\le 1$ and
$1-\varepsilon^2(1-\varepsilon)^2\le\gamma\le 1$, it is immediate
to see that the system state never reaches the maximally mixed
state and that $\max \delta_{12}(t)= 2/3$, which is well below the
limit of 8/9 given for bipartite qubit states. Although the state
is separable, as we will shortly discuss.

 An important measure of entanglement is the negativity of the
state calculated as \begin{math}(C^{2}\otimes C^{2})\end{math}
\cite{{TzuChieh03},{Verstraete01}}. This last criterion is related
to the separability of the state considering that the state is
separable if the partially  transposed state is also a valid
quantum state, that is a positive semidefinite operator
\cite{{TzuChieh03},{Verstraete01}}. The partial transposition of a
non-separable state presents one negative eigenvalue and thus we
need to follow the eigenvalues of the partially transposed joint state.
For the calculation of the negativity we have considered the definition
\cite{{TzuChieh03},{Verstraete01}}
\begin{eqnarray}
    N(\rho,t)=2\max\left\{0,-\lambda_{neg}(t)\right\}.
\end{eqnarray}
For the initial state (\ref{inst}) here considered
\begin{eqnarray}
N(\rho_\varepsilon,t)&=&\frac{2\varepsilon(1-\varepsilon)|A|}{\varepsilon^2+(1-\varepsilon)^2}\nonumber\\
&=&\frac{2\varepsilon(1-\varepsilon)\sqrt{1-\gamma}}{\varepsilon^2+(1-\varepsilon)^2}.\end{eqnarray}
Notice that for $t=0$
\begin{eqnarray}
N(\rho_\varepsilon,0)&=&\frac{2\varepsilon(1-\varepsilon)}{\varepsilon^2+(1-\varepsilon)^2},\end{eqnarray}
which is maximal for $\varepsilon=1/2$. $N(\rho_\varepsilon,0)$ is
exactly the coherence of the initial pure state given by
(\ref{inst}) and its maximal value represents the maximally
entangled pure state given by (\ref{inst})
 for $\varepsilon=1/2$. For this special case
\begin{eqnarray}
    N(\rho_{1/2},t)=e^{\widetilde{N}[\cos((\mu_{1}+\mu_{2})t)-1]}=\sqrt{1-\gamma}.
\end{eqnarray}
Although $N(\rho_{1/2},t)$ does not change sign it gets rapidly
closer to zero for $\widetilde{N}\gg1$. Only for
$N(\rho_{1/2},t)=0$ (or $\gamma=1$) the system is {\sl separable}.

 Another important measure of entanglement which has an exact
 analytic expression for coupled qubits is the entanglement of formation
\cite{{Munro01},{Verstraete01},{Hill97}}. It is defined as
 \begin{eqnarray}
E_{F}=h\left(\frac{1}{2}\left[1+\sqrt{1-C(\rho)^{2}}\right]\right).
\end{eqnarray} being $h$ and the concurrence
\begin{math}C(\rho)\end{math} defined as
\cite{{Munro01},{Verstraete01},{Hill97}}
\begin{eqnarray}
h_{x}=-x\log_{2}x-(1-x)\log_{2}(1-x),
\end{eqnarray}
\be C(\rho)\equiv
\max\{0,\sqrt{\lambda_1}-\sqrt{\lambda_2}-\sqrt{\lambda_3}-\sqrt{\lambda_4}\},
\ee where $\lambda_i$ are the eigenvalues of
$\rho\sigma_2\otimes\sigma_2\rho^*\sigma_2\otimes\sigma_2$, {
where} $\sigma_2$ { is} the Pauli $\sigma_y$-spin matrix. For the
above state (\ref{inst}),
\begin{eqnarray}
C(\rho_\varepsilon)&=&\sqrt{1-\frac{2\varepsilon^2(1-\varepsilon)^2(1-|A|^2)}{[\varepsilon^2+(1-\varepsilon)^2]^2}.}
\end{eqnarray}
For the maximally entangled state
\begin{eqnarray}
C(\rho_{1/2})&=&\sqrt{\frac{1}{2}\left\{1+e^{2\widetilde{N}\left[cos((\mu_{1}+
\mu_{2})t)-1\right]}\right\}}\nonumber\\
&=&\sqrt{1-\frac{\gamma}{2}}
\end{eqnarray}
For the special dynamical system considered the above mentioned
measures are all {monotonic} functions of each other. For
example we can write all the other measures as a function of the
negativity $N(t)$ as
\begin{eqnarray}
\delta_{12}(\varepsilon,t)&=&\frac23\left(N^2(\rho_\varepsilon,0)-N^2(\rho_\varepsilon,t)\right),\\
C(\rho_\varepsilon)&=&\sqrt{1-\frac12\left(N^2(\rho_\varepsilon,0)-N^2(\rho_\varepsilon,t)\right)},
\end{eqnarray} which for $\varepsilon=1/2$ writes
\begin{eqnarray}
\delta_{12}(1/2,t)=\frac{2}{3}\left(1-N(\rho_{1/2},t)^2\right)\\
C(\rho_{1/2})=\sqrt{\frac{1}{2}\left(1+N(\rho_{1/2},t)^{2}\right)}.
\end{eqnarray}

\section{Distance as entanglement measures}
While the above considered entanglement measures are capable to
quantify the amount of entanglement present in the quantum state
(\ref{state}) they lack an interpretative meaning.
 It is possible to define an entanglement measure ${\cal E}
(\rho,\sigma)$ of a quantum state $\rho$ as a distance measure
between the quantum state $\rho$ and a reference  state $\sigma$.
The distance must be minimized over all the dynamically connected
reference states $\sigma$. For example, the Bures distance
\cite{bures} was recently identified as a possible quantification
of entanglement \cite{vedral,paulina2,entclas}, for one and two
parties states, respectively. The Bures distance is defined as \be
\label{bures}
d_B(\rho,\sigma)=(2-2\sqrt{\mathcal{F}(\rho,\sigma)})^{1/2},\ee
where ${\mathcal{F}}(\rho,\sigma)$ is the Uhlmann Fidelity
\cite{bures} between any two quantum states $\rho$ and $\sigma$:
${\mathcal{F}}(\rho,\sigma)=\{Tr[(\sqrt{\rho}\sigma\sqrt{\rho})^{1/2}]\}^2$,
ranging from 0 to 1. In such a case $d_B(\rho,\sigma)$ must be
minimized over the set the possible referential $\sigma$ states.

Two choices of the reference state can be made, from which the
distance measure definition as an entanglement measure will be
dependent: (i) a mixed state reference, and (ii) a pure state
reference. A mixed state reference is a natural choice, since it
 was proven that a pure reference state does not allows that the
 distance-based entanglement measure
be an entanglement monotone (see \cite{vedral,paulina2}), once it
can always be increased by appropriate local operations on $\rho$.
On the other hand, the redundancy of possible mixed reference
states, as we will discuss in what follows, and the appealing
physical meaning that a pure reference state offers, make it
interesting for comparison with the previously described
entanglement measures. In what follows we will consider both (i)
and (ii) situations, and show how do they relate to each other.

 \subsection{Mixed reference state}

For a mixed reference state, the closer $\rho_\varepsilon$ is from
the reference $\sigma$, the less pure it will be and thus the
state will be less entangled. That means that for a mixed
reference state { $\sigma_m$,} the Bures distance itself can be
regarded as an entanglement measure \cite{paulina2} \be
{\mathcal{E}}(\rho_\varepsilon,\sigma_m)=\min_{\sigma\in
{\mathcal{D}}}\frac{1}{2}{d_B(\rho_\varepsilon,\sigma_m)}^2, \ee
where $\mathcal{D}$ is the set of all separable bipartite states
of the system. ${\mathcal{E}}(\rho_\varepsilon,\sigma_m)$ was
numerically calculated and it will be presented in next section.
For the maximally entangled initial state considered here
($\varepsilon=1/2$), $ {\mathcal{E}}(\rho_{1/2},\sigma_m)$ has a
simple expression. In this situation the reference state which
gives the minimal distance is the following state:
\begin{eqnarray}
\sigma_{m}=\left( \begin{array}{cccc} \frac{1}{2} & 0 & 0 &
\frac{e^{-i(\omega_{1}+\omega_{2})t}}{2 e^{2\widetilde{N}}
}\\ 0 & 0 & 0 & 0\\ 0 & 0 & 0 & 0\\
\frac{e^{i(\omega_{1}+\omega_{2})t}}{2 e^{2\widetilde{N}} } & 0 &
0 & \frac{1}{2} \end{array} \right).
\end{eqnarray}
For this state  the fidelity is
\begin{eqnarray}
{\mathcal{F}}(\rho_{1/2},\sigma_{m})=\frac{1}{4}\left[\sqrt{1+\beta-\alpha}+
\sqrt{1+\beta+\alpha}\right]^2,
\end{eqnarray}
and the entanglement measure will be \be
{\mathcal{E}}(\rho_{1/2},\sigma_m)= 1-
\frac{1}{2}\left[\sqrt{1+\beta-\alpha}+\sqrt{1+\beta+\alpha}\right],\ee
with
\begin{eqnarray}
\alpha=\sqrt{(-1-\beta)^{2}-(-1+\left|A\right|^{2})(-1+4\left|X\right|^{2})}
\end{eqnarray}
\begin{eqnarray}
\beta=A^{*}X+AX^{*},
\end{eqnarray} and
\begin{eqnarray}
X={e^{-i(\omega_{1}+\omega_{2})t}}{e^{-2\widetilde{N}} }
\end{eqnarray}

As such the entanglement measure ranges from $0$ for a separable
state to $1-\frac{1}{\sqrt{2}}\sqrt{1+e^{-2\widetilde{N}}}$ for a
maximally entangled state.

Notice that a natural choice would be to use as a reference
separable state the maximally mixed state
\begin{eqnarray} \label{maxmix}
    \sigma_{mm}=\frac{I}{4}.
\end{eqnarray}
However, such a state does not belong to the set of separable
states attained by the quantum system here considered in the
chosen conditions. We will consider this state for comparison with
the result obtained for an actual state $\sigma$ belonging to
$\mathcal{D}$ and show that, despite their distinction, the
distance measures introduced above approximate each other as
$\widetilde{N}\rightarrow\infty$. {With respect} to the {\sl
maximally mixed state} (\ref{maxmix}) the fidelity  of the quantum
state gives
\begin{eqnarray}
{\mathcal{F}}
(\rho_{1/2},\sigma_{mm})=\frac{1}{4}\left[\sqrt{1-{\left|A\right|}}
+\sqrt{1+{\left|A\right|}}\right]^{2},
\end{eqnarray}
for $A$ given by (\ref{a}). The entanglement measure gives \be
{\mathcal{E}}(\rho_{1/2},\sigma_{mm})= 1-
\frac{1}{2}\left[\sqrt{1-{\left|A\right|}}+\sqrt{1+{\left|A\right|}}
\right],\ee thus ranging from {zero to }
$1-\frac{1}{2}\left(\sqrt{1-
e^{-2\widetilde{N}}}+\sqrt{1+e^{-2\widetilde{N}}}\right)$ for a
separable state, and {from zero to } $1-\frac{1}{\sqrt{2}}$ for a
maximally entangled state. Notice that as
$\widetilde{N}\rightarrow\infty$, then
${\mathcal{E}}(\rho,\sigma_{mm})\rightarrow{\mathcal{E}}(\rho,\sigma_{m})$,
despite being for different reference states.

\subsection{Pure reference state}

For a pure reference state the Bures distance measure must
 be minimized over all the dynamically connected reference pure
 states, thus satisfying the following criteria
\begin{eqnarray}
Tr_{1}\left\{\rho^P_{12}(t)\right\}\equiv\rho_{2}(t)\nonumber\\
Tr_{2}\left\{\rho^P_{12}(t)\right\}\equiv\rho_{1}(t),
\end{eqnarray}
and the Uhlmann fidelity  reduces to the usual fidelity,
${\mathcal{F}}(\rho(t),\sigma_p)=Tr\{{\rho(t)}\sigma_p\}$. Now the
Bures distance writes simply as
 \be \label{bures2}
d_B(\rho,\sigma)=(2-2{\mathcal{F}}(\rho,\sigma))^{1/2}.
\ee
Remark that the relative entropy related to the pure state $\sigma_p$
is defined as
\br
E_r(\rho,\sigma_p)&=&Tr\left\{\sigma_p\log_2\sigma_p-\sigma_p\log_2
\rho(t)\right\}\nonumber\\
&=&-Tr\left\{\sigma_p\log_2\rho(t)\right\},\er
in binary units or
\be E_r(\rho,\sigma_p)=-Tr\left\{\sigma_p\ln\rho(t)\right\},
\ee
in natural units. Since $\rho(t)=1-(1-\rho(t))$, such that $ Tr
\{1-\rho(t)\}<1$, thus
$\ln[1-(1-\rho(t))]=-[(1-\rho(t))+(1-\rho(t))^2/2+...]$ and the
relative entropy can be written to first order in $(1-\rho(t))$
as
\br E_r(\rho,\sigma_p)&\approx&
1-Tr\left\{\sigma_p\rho(t)\right\}\nonumber\\&=&1-{\mathcal{F}}
(\rho,\sigma_p).
\er
That means that the Uhlmann fidelity of the quantum state
$\rho(t)$ to a pure reference state $\sigma_p$ corresponds to one
minus the linearized relative entropy. Now defining the
entanglement measure as \be
{\mathcal{E}}(\rho,\sigma_p)=1-\min_{\sigma_p\in
{\mathcal{D}}}\frac{1}{2}{d_B(\rho,\sigma_p)}^2,\ee we obtain \be
{\mathcal{E}}(\rho,\sigma_p)={\mathcal{F}}(\rho,\sigma_p)=1-E_r(\rho,\sigma_p),\ee
which shows in a nice way how the distance entanglement measure
relates to the relative entropy for pure reference states.

The reference state is simply a purified version of the studied
quantum state. It could in fact represent a whole family of
states, if we have not considered that the system dynamics
restricts the possible reference states as follows. {Since} the
system dynamics does not allow energy transference between
subsystems the initial unpopulated subspace
\begin{math}(|+,-\rangle,|-,+\rangle)\end{math} does not participate
in the choice of the pure reference state. Moreover, the intrinsic
system dynamics must be included in the reference state. The
perfect choice is $\sigma_p=\rho_{\varepsilon}^{0}$ given by Eq.
(\ref{rho0}).

 Thus in this case the fidelity of the system state to the pure
 reference state (\begin{math}\sigma_{p}\end{math}) and thus the
 entanglement measure writes as
\br
{\mathcal{E}}(\rho_\varepsilon,\sigma_p)&=&\frac{\left[\varepsilon^4+(1-\varepsilon)^4+2\varepsilon^2(1-\varepsilon)^2|A|\right]}{\left[\varepsilon^4+(1-\varepsilon)^4+2\varepsilon^2(1-\varepsilon)^2\right]}\nonumber\\
&=&1-\frac{2\varepsilon^2(1-\varepsilon)^2(1-|A|)}{[\varepsilon^2+(1-\varepsilon)^2]^2},
\er or \br
{\mathcal{E}}(\rho_\varepsilon,\sigma_p)&=&1-\frac12N(\rho_\varepsilon,0)\left(N(\rho_\varepsilon,0)-N(\rho_\varepsilon,t)\right).\er
For $\varepsilon=1/2$ \br
{\mathcal{E}}(\rho_{1/2},\sigma_p)&=&\frac{1}{2}\left\{1+e^{\widetilde{N}
[\cos((\mu_{1}+\mu_{2})t)-1]}\right\}\nonumber\\&=&\frac{1}{2}\left\{1+
N(\rho_{1/2},t)\right\}, \er thus ranging from $\frac 1
{\sqrt{2}}\sqrt{1+e^{-2\widetilde{N}}}$ to 1 for a separable state
and a maximally entangled state, respectively. Notice that this
range is displaced in $\frac {1}{\sqrt{2}}
\sqrt{1+e^{-2\widetilde{N}}}$ in relation to the range for
${\mathcal{E}} (\rho,\sigma_m) $.

 As such the relative entropy
writes as \br
E_r(\rho_{\varepsilon},\sigma_p)=\frac12N(\rho_\varepsilon,0)\left(N(\rho_\varepsilon,0)-N(\rho_\varepsilon,t)\right)\er
and
 \be E_r(\rho_{1/2},\sigma_p)=\frac{1}{2}\left(1-N(t)\right).
\ee

We remark that for pure states $E_F=E_r$, and for a general mixed
state $E_F\ge E_r$.

\section{Entanglement measures ordering}
Before we compare the entanglement measures by varying the degree
of mixing we observe that the state purity and the concurrence can
be related to the linearized relative entropy and the distance
entanglement measure as
\begin{widetext}
\br
\delta_{12}(\rho_{\varepsilon},t)N^2(\rho_\varepsilon,0)&=&\frac{2d}{d-1}
\left\{E_r(\rho_{\varepsilon},\sigma_p)
\frac{2d}{d-1}
\left({\mathcal{E}}(\rho_{\varepsilon},\sigma_p)-1\right)\right.%\nonumber\\
%&&
+\frac12N(\rho_\varepsilon,0)\left[E_r(\rho_{\varepsilon},\sigma_p)
\left(1+N(\rho_\varepsilon,0)\right)\right.\nonumber\\
&&
\left.\left.+\left({\mathcal{E}}(\rho_{\varepsilon},\sigma_p)-1\right)
\left(1-N(\rho_\varepsilon,0)\right)\right]\right\}\\
C(\rho_{\varepsilon})^2N^2(\rho_\varepsilon,0)
&=&
\left({\mathcal{E}}(\rho_{\varepsilon},\sigma_p)-1\right)^2
+E_r^2(\rho_{\varepsilon},\sigma_p)%\nonumber\\
%&&
+{\mathcal{E}}(\rho_{\varepsilon},\sigma_p)N(\rho_\varepsilon,0)\left(1+N(\rho_\varepsilon,0)\right)\nonumber\\
&&
+E_r(\rho_{\varepsilon},\sigma_p)N(\rho_\varepsilon,0)\left(1-N(\rho_\varepsilon,0)\right),
\er
\end{widetext}
 Notice that for the maximally entangled state at
$\varepsilon=1/2$ the state purity can be related to the
linearized relative entropy and the distance entanglement measure
as \br
\delta_{12}(\rho_{1/2},t)&=&\frac{2d}{d-1}{\mathcal{E}}(\rho_{1/2},\sigma_p)
E_r(\rho_{1/2},\sigma_p), \er while the concurrence can be written
as \br C(\rho_{1/2})^2&=&{\mathcal{E}}^2(\rho_{1/2},\sigma_p) +
E_r^2(\rho_{1/2},\sigma_p), \er thus defining a quite interesting
triangular relation between the concurrence the distance
entanglement measure and the relative entropy.
%FIGURE 1

The relation between entanglement and mixing measures can be used
to define an entanglement measure ordering against the degree of
mixing. In Fig. 1 we plot $N(\rho_\varepsilon)$,
$C(\rho_\varepsilon)$,
${\mathcal{E}}(\rho_{\varepsilon},\sigma_m)$, and
${\mathcal{F}}(\rho_\varepsilon,\sigma_p)$ as given by Eqs. (18,
24, 31, and 48), respectively, against the degree of mixing
(linear entropy), Eq. (16), for the family of states determined by
$\varepsilon$. An important feature is that for each
$\varepsilon$, there is no crossing between the measures, which
implies that a well-defined ordering can be given for the
respective state. Thus, let $\rho_\varepsilon^0$ be the initial
quantum state under a phase damping channel. Then always
${\mathcal{E}}(\rho_{\varepsilon},\sigma_m)\le
N(\rho_\varepsilon)\le{\mathcal{F}}(\rho_\varepsilon,\sigma_p)\le
C(\rho_\varepsilon)$ for a given $\varepsilon$. On the other hand
let us consider the whole family of states $\rho_\varepsilon(t)$
for a given linear entropy. Then always
$N(\rho_\varepsilon)\le{\mathcal{F}}(\rho_{\varepsilon^\prime},\sigma_p)\le
C(\rho_{\varepsilon^{\prime\prime}})$ for any $\varepsilon$,
$\varepsilon^\prime$, and $\varepsilon^{\prime\prime}$. But for a
given degree of mixing, there is no obvious ordering between
${\mathcal{E}}(\rho_{\varepsilon},\sigma_m)$ and $
N(\rho_{\varepsilon^\prime})$ (see the discussion bellow however).
Now in Fig. 2 we compare the entanglement measures together with
the linear entropy against the fidelity to the pure state,
${\mathcal{F}}(\rho_{\varepsilon^\prime},\sigma_p)$. Here for a
given $\varepsilon$ it is observed
${\mathcal{E}}(\rho_{\varepsilon},\sigma_m)\le
N(\rho_\varepsilon)\le C(\rho_\varepsilon)$. But for a fixed
fidelity no obvious ordering between
${\mathcal{E}}(\rho_{\varepsilon},\sigma_m)$ and $
N(\rho_{\varepsilon^\prime})$ is observed. In fact the only
possible relation is that for a given fidelity (or linear entropy)
then ${\mathcal{E}}(\rho_{\varepsilon},\sigma_m)\le
N(\rho_{\varepsilon^\prime})$, for
$\varepsilon\le\varepsilon^\prime$, which is however a very weak
relation since there are many $\varepsilon>\varepsilon^\prime$
that also satisfy this inequality. Thus no ordering can be
identified for ${\mathcal{E}}(\rho_{\varepsilon},\sigma_m)$ and
$N(\rho_{\varepsilon^\prime})$.

%However, the essential point is that for
%$\varepsilon\neq\varepsilon^\prime$ the quantum states
%$\rho_{\varepsilon}$ and $\rho_{\varepsilon^\prime}$ are different
%and the ordering of entanglement measures can be different, since
%the computed measures are strongly dependent on both the initial
%sets of quantum states and the reference quantum states. Such a
%measures ordering for $\varepsilon=
%\varepsilon^\prime$ can be necessary and sufficient condition (for what?).}\\

\section{Concluding Remarks}

The present work was motivated by the need to analyze the
 degree of entanglement against mixing for a specific dynamical system composed
of two coupled qubits under a phase damping channel. We have
discussed the ordering of some possible entanglement measures for
a family of pure initial states. We have considered as
entanglement measures the negativity $N$ and concurrence
(entanglement of formation) $C$, that is calculated analytically,
and the Fidelity $\mathcal{F}$ and  Bures distance entanglement
measure $\mathcal{E}$, which is calculated numerically and
analytically for some case. For the fidelity $\mathcal{F}$ and
Bures distance entanglement measure $\mathcal{E}$ it is necessary
to define a reference state to compute the measures. The results
have shown that the dynamics of the system restricts the
possibilities and determine the reference pure state associated to
the specific dynamics of the system. In the case of a mixed
reference state, we have discussed that the maximally mixed state
is not the best choice. Instead it is necessary to choose a mixed
state associated to the dynamics of the system. Then the model
with phase damping channel has suggested that the best reference
states are always associated to the dynamics of the system.

We have shown an entanglement measures ordering for a family of
initial states.  As the considered entanglement measures are
strongly dependent on the initial state and reference state, the
measures ordering was then determined for definite values of $
\varepsilon$. These results have shown that the Bures distance can
be envisaged as a possible quantitative and qualitative measure of
entanglement.

\acknowledgments{
%%%%%%%%%%%%%%%%%%%%%%%%%%%%%%%%%%%%%%%%%%%%%%%%%%%%%%%%%%%%%
MCO acknowledges funding in part by FAPESP under project
$\#04/14605-2$ and by FAEPEX-UNICAMP. ESC and KF  acknowledge
support from CNPq under projects $\#140243/2001-1$ and
$\#300651/85-6$.}

%\vspace*{-0.5cm}

%FIGURE 1

%\end{multicols}
\newpage
{\bf Figure Captions}

\vspace{1cm}
 Fig. 1 - Entanglement measures against linear entropy
for a range of $\rho_\varepsilon^0$ initial states.
$\varepsilon=0.5$ (solid line), $\varepsilon=0.42$ (dashed line),
$\varepsilon=0.37$ (dotted line), $\varepsilon=0.32$ (dash-dotted
line), $\varepsilon=0.26$ (short-dashed line), $\varepsilon=0.19$
(short-dotted line), $\varepsilon=0.12$ (short-dash-dotted line),
and $\varepsilon=0.05$ (dash-dotted-dotted line).

\vspace{1cm}
 Fig. 2 - Entanglement and mixing measures against
pure state fidelity for a range of $\rho_\varepsilon^0$ initial
states. $\varepsilon=0.5$ (solid line), $\varepsilon=0.42$ (dashed
line), $\varepsilon=0.37$ (dotted line), $\varepsilon=0.32$
(dash-dotted line), $\varepsilon=0.26$ (short-dashed line),
$\varepsilon=0.19$ (short-dotted line), $\varepsilon=0.12$
(short-dash-dotted line), and $\varepsilon=0.05$
(dash-dotted-dotted line)

\end{document}